\documentclass[letterpaper]{article} 
\usepackage{aaai25}  
\usepackage{times}  
\usepackage{helvet}  
\usepackage{courier}  
\usepackage[hyphens]{url}  
\usepackage{graphicx} 
\usepackage{amsmath}
\usepackage{accessibility}
\usepackage{rotating}
\usepackage{natbib}  
\usepackage{caption} 
\usepackage{algorithm}
\usepackage{algorithmic}

\usepackage{newfloat}
\usepackage{listings}
\DeclareCaptionStyle{ruled}{labelfont=normalfont,labelsep=colon,strut=off} 
\floatstyle{ruled}
\newfloat{listing}{tb}{lst}{}
\floatname{listing}{Listing}
\nocopyright 

\title{Style4Rec: Enhancing Transformer-based E-commerce Recommendation Systems with Style and Shopping Cart Information}
\author{
    Berke Ugurlu\textsuperscript{\rm 1},
    Ming-Yi Hong\textsuperscript{\rm 1},
    Che Lin\textsuperscript{\rm 1}
}
\affiliations{
    \textsuperscript{\rm 1}National Taiwan University\\

    Taipei, Taiwan
}

\usepackage{bibentry}

\begin{document}

\maketitle

\begin{abstract}
Understanding users' product preferences is essential to the efficacy of a recommendation system. Precision marketing leverages users' historical data to discern these preferences and recommends products that align with them. However, recent browsing and purchase records might better reflect current purchasing inclinations. Transformer-based recommendation systems have made strides in sequential recommendation tasks, but they often fall short in utilizing product image style information and shopping cart data effectively. In light of this, we propose Style4Rec, a transformer-based e-commerce recommendation system that harnesses style and shopping cart information to enhance existing transformer-based sequential product recommendation systems. Style4Rec represents a significant step forward in personalized e-commerce recommendations, outperforming benchmarks across various evaluation metrics. Style4Rec resulted in notable improvements: HR@5 increased from 0.681 to 0.735, NDCG@5 increased from 0.594 to 0.674, and MRR@5 increased from 0.559 to 0.654. We tested our model using an e-commerce dataset from our partnering company and found that it exceeded established transformer-based sequential recommendation benchmarks across various evaluation metrics. Thus, Style4Rec presents a significant step forward in personalized e-commerce recommendation systems.
\end{abstract}

%

\section{Introduction}
Sequential product recommendation is a process that involves capturing information from user sessions to predict the next item they are likely to purchase. By analyzing the history of user interactions, the system aims to make accurate predictions. However, due to the intricate relationships between products, effectively capturing information from users' past interactions poses a significant challenge. 

Different approaches have been explored in the realm of sequential product recommendation to capture the dependencies between products. One simple model is Markov Chains (MCs) \cite{MC}, which predict the subsequent product based on the previous product or a few preceding products. However, MCs do not effectively utilize long-range dependencies and focus primarily on the last few products, attempting to capture short-range product dependencies. On the other hand, Recurrent Neural Networks (RNNs) \cite{hidasi2016sessionbased} have been employed to capture both long and short-term dependencies between products. By utilizing a hidden state, RNNs are able to predict the subsequent product. GRU4Rec \cite{hidasi2016sessionbased}, and its improved versions have demonstrated success in the task of next-item prediction. GRU4Rec takes both the hidden state and current product vector as input, leveraging their combination for accurate predictions. A different approach is adopted in CNN-based sequential product recommendation models \cite{caser}. Here, the embeddings of previous products are treated as an image, and convolutional operations are applied to extract relevant information for predicting the subsequent product. This approach utilizes the structural properties of CNNs to capture dependencies between products and make effective predictions.

Transformer-based sequential product recommendation models have emerged as promising alternatives \cite{bert4rec,sasrec,ssept} in predicting subsequent items in user sessions, showcasing state-of-the-art performance. Unlike previous algorithms, transformer-based models leverage self-attention mechanisms, which are highly efficient in training and excel at extracting patterns within user sessions. The self-attention mechanism allows the model to weigh the importance of different elements in the input sequence, capturing dependencies and relationships effectively. As a result, transformer-based models have proven to be powerful tools for sequential product recommendations, outperforming traditional approaches in terms of accuracy and efficiency.

Existing transformer-based sequential models \cite{attrec,fissa,bert4rec,sasrec,ssept} utilizes only the purchase information for predicting the next item in user sessions. Product images undoubtedly play a major role in users' preferences. However, existing transformer-based sequential product recommendation models don't have a methodology to incorporate the style information of the product images. Utilizing style information can greatly improve the performance in sequential recommendation tasks and allow us to evaluate user preferences more thoroughly.

In our research, we have developed a multi-layer transformer-based sequential product recommendation system that leverages style information and shopping cart data to enhance existing state-of-the-art recommendation systems. By grouping users based on their individual sessions, we could recommend multiple products for each user. The performance of our system surpassed that of previous recommendation models designed for sequential product recommendation tasks, including Bert4Rec \cite{bert4rec} and SASRec \cite{sasrec}, both of which are transformer-based. Our results demonstrate the effectiveness and superiority of our approach in providing personalized and accurate product recommendations.  

In our sequential product recommendation model, we utilized a dataset provided by our partnering company, encompassing user history spanning a significant period of 1.5 years. The dataset included valuable information such as purchase data, shopping cart data, and product images.
Recognizing the crucial role of product images in influencing user preferences, we employed the neural style transfer algorithm \cite{style} to extract style information from the product images. This style information was then used to create style embeddings, enhancing our recommendation system's performance. Our decision to extract style information was driven by the fact that the product category was already known, and each product image represented a single item. Consequently, directly utilizing an object detection algorithm layer would not provide us with significant insights, as such layers focus on capturing category information that is already known. Instead, we leveraged the correlation of feature maps in those layers by using the VGG-19 object detection algorithm \cite{vgg} to extract the style information, as described in the neural style transfer algorithm \cite{style}. This modified approach enabled us to create style embeddings specifically tailored for our multi-layer transformer-based sequential recommendation network.

Furthermore, we incorporated the shopping cart data of the users into our recommendation system. Recognizing that products added to the shopping cart reflect user interest, even if they were not ultimately purchased, we developed a training strategy to account for the distinction between purchase and shopping cart products. The specific utilization of shopping cart data and the extraction of style embeddings within our sequential recommendation system are thoroughly evaluated and detailed in the methodology section.

Existing sequential product recommendation systems, such as Bert4Rec \cite{bert4rec}, SaSRec \cite{sasrec}, and SSE-PT \cite{ssept}, have demonstrated remarkable success in predicting subsequent items for user sessions. However, these models solely rely on purchase data as their input for the transformer network. This limitation prevents them from effectively utilizing valuable information present in the product images and shopping cart data available in the e-commerce dataset provided by our partnering company.

To address this gap, we propose two methods to enhance the performance of sequential product recommendation systems. Firstly, we leverage the neural style transfer algorithm \cite{style} to extract style information from product images, which is then utilized as style embeddings in our model. This allows us to incorporate important visual cues into the recommendation process. Secondly, we adopt a strategy where shopping cart data is employed exclusively during the training and validation phases and excluded during testing. This approach provides a more accurate evaluation of real-world performance.

Through extensive experiments conducted on our dataset, our proposed model has surpassed the existing state-of-the-art baselines, showcasing its effectiveness in improving sequential product recommendations. These findings validate the significance of incorporating style information from product images and leveraging shopping cart data to enhance the performance of recommendation systems in the e-commerce domain.

The main contributions of our research are as follows:
\begin{itemize}
\item We have designed and implemented a novel transformer-based model for sequential product recommendation. This model incorporates separate components for obtaining the product vector of user history and the learnable product vector. 
\item We proposed a style extraction module that effectively obtains style embeddings from product images using the neural style transfer algorithm. This algorithm was modified to ensure its compatibility with transformer-based sequential product recommendation networks. 
\item We developed a method to differentiate between purchase and shopping cart sessions. Specifically, we employed shopping cart sessions exclusively during the training and validation phases while excluding them during testing. This approach allowed us to effectively capture the distinction between these types of sessions and incorporate it into our recommendation model.
\end{itemize}

These contributions highlight the unique features and advancements of our transformer-based sequential product recommendation network, emphasizing its ability to effectively combine different types of embeddings for improved recommendation performance.

\section{Related Work}

AttRec \cite{attrec}, SASRec \cite{sasrec}, SSE-PT \cite{ssept}, FISSA \cite{fissa}, and BERT4Rec \cite{bert4rec} are among the notable transformer-based models used for sequential product recommendation tasks. These models employ multi-layer transformer blocks to capture item-item relations within user sessions.

AttRec \cite{attrec} leverages the self-attention mechanism to capture both long-term and short-term interactions between items in user sessions. It considers the temporal dynamics of the interactions separately. SASRec \cite{sasrec} utilizes multiple transformer blocks that facilitate left-to-right item-item interactions. It truncates user sessions and performs separate predictions for each truncated session, allowing the model to capture sequential patterns effectively. SSE-PT \cite{ssept} extends SASRec by incorporating personalized user embeddings and employs the stochastic shared embedding (SSE) regularization technique to mitigate overfitting. FISSA \cite{fissa} introduces a global representation learning module, a local representation learning module, and a gating module to balance the impact of global and local representations using a multi-layer perceptron (MLP) layer. BERT4Rec \cite{bert4rec} adopts a bidirectional training framework using the \(Cloze\) task. It employs input masking during training, arguing that unidirectional transformer architectures may limit the true potential of transformer-based recommendation systems.

The aforementioned transformer-based sequential recommendation models have demonstrated remarkable effectiveness in sequential product recommendation tasks. However, none of these models have addressed the utilization of product images and shopping cart data to enhance performance. It is intuitive that product images play a significant role in shaping users' preferences, while shopping cart products offer valuable insights as they represent items of interest to users.

In our transformer-based sequential recommendation model, we addressed the incorporation of product images and shopping cart data by introducing multiple modules. One of the modules involved obtaining the product vector of historical behavior, which captured the user's interactions with products over time. This vector represented the user's preferences and interests based on their past choices. We used the neural style transfer algorithm \cite{style} to extract style for utilizing the product image embeddings. This allowed us to incorporate visual style information into the recommendation process, enhancing the model's ability to capture visual product attributes. 

In summary, our transformer-based sequential recommendation models incorporated product images and shopping cart data by introducing multiple modules. This included obtaining the product vector of historical behavior, utilizing learnable product vectors, and incorporating style embeddings extracted through the neural style transfer algorithm \cite{style}. The neural style transfer algorithm \cite{style}, wasn't utilized in existing transformer-based sequential product recommendation systems. Our work showcases the first model incorporating style information into transformer-based sequential product recommendation models. Regarding the temporal aspect, we did not consider time-aware sequential product recommendation models due to the lack of precise time information for purchased products. We leveraged the order of the products in constructing user sessions to capture sequential patterns.

\subsection{Problem Statement}
Each user session can be considered as sequential data and sequential product recommendation systems try to predict the next item that the user might buy for user sessions.
In sequential product recommendation systems, given a set of users and set of items, \(U = \{u_1, u_2, u_3,.....,u_{|U|}\} \) and \(I = \{i_1, i_2, i_3,.....,i_{|I|}\} \),  we can construct user sessions as \(S_u = [i_1^u, i_2^u,...,i_t^u,...,i_{n_u}^u] \), $u\in U$ in chronological order. 
The length of the session is \(n_u\), and \(i_t\) is the product that the user interacted with at time t. The purpose of the sequential product recommendation is to predict which item the user will interact with at time t+1, given the user's history \(S_u\). For the final prediction, the products on the product list \(I\) are sorted with respect to their relevance scores.

\section{Methodology}

Our multi-layer transformer-based sequential recommendation network consists of two parts, the first part utilizes the deep transformer encoder, and the second part utilizes the modified version of the neural style transfer algorithm \cite{style}. The second part was used for extracting the style information from the product images, and the style information was used for further performance increases.

\subsection{Deep Transformer Encoder}

The deep-transformer-encoder network consists of multiple transformer-encoder blocks on top of each other. The hidden representation of the input was calculated for each transformer-encoder block and fed into the next transformer-encoder block. We utilized the multi-head self-attention mechanism and point-wise feed-forward network for constructing the transformer-encoder blocks, \cite{attention,bert4rec}: 

\begin{align}
    multiHead(H^l) &= [head_1;head_2,....,head_h]W^o\\
    head_i &= Att(H^lW_i^Q, H^lW_i^K, H^lW_i^V)\\
    Att(Q, K, V) &= Softmax(\frac{QK^T}{\sqrt{d/h}})V
\end{align}

The formulas (1), (2), and (3) show the details of the multi-head attention mechanism. \(H^l\) represents the stacked hidden representation for a given product sequence. \(H^l\) is projected into different subspaces by using the key, query, and value matrices, which are \(W^Q\), \(W^K\), \(W^V\). The subscript \(i\) represents the multi-head index. The heads are concatenated and projected by utilizing \(W^O\) matrix. \(W^Q\), \(W^K\), \(W^V\), \(W^O\) are learnable projection matrices, \cite{attention,sasrec,bert4rec,li2018multihead}. We used scaled-dot product attention, \cite{attention,hinton2015distilling}. The query Q, the key K, and the value V are projected from stacked hidden representation,  \(H^l\). Then, we applied a point-wise feed-forward network to the result of the multi-head self-attention in order to utilize non-linearity and the interactions between different dimensions. We also utilized residual connection, layer normalization, and dropout to avoid overfitting the data \cite{ba2016layer,residual,overfitting}. 

As you can see in Figure \ref{model}, to calculate the binary cross-entropy loss, we first extracted the product vector of historical behavior for each session. We then compared this vector with both the learnable product embedding of the ground truth product and that of the negatively sampled product. Cosine similarity was utilized to compute the relevance score of these vectors, which was then converted into probabilities using the softmax function. These probabilities were subsequently used to calculate the binary cross-entropy loss.

The fundamental rationale behind this model architecture is to enhance the cosine similarity between the product vector of historical behavior and the ground truth vector. Simultaneously, we aim to decrease the relevance, in terms of cosine similarity, between the product vector of the historical behavior and the vector of the negatively sampled product. This dual approach optimizes the model's ability to discern relevant and irrelevant products based on historical behavior.


\subsection{Embedding Extraction Module}

The embedding extraction module in Figure 3 consists of 2 parts, which are:
\begin{enumerate}
    \item Sinusoidal Positional Embeddings
    \item Learnable Product Embeddings
\end{enumerate}

We employed a combination of sinusoidal positional embeddings and learnable product embeddings. Our goal was twofold: firstly, to encode information about the relative positions of products within sessions, and secondly, to generate unique learnable embeddings for each product, enabling their differentiation and comparison.

Sinusoidal positional embeddings were utilized to encode the products' relative positions within sessions, while learnable product embeddings facilitated product comparison, thus informing product recommendations. This approach integrated both positional information and product-specific embeddings simultaneously.

 
\begin{align}
    PE_{(pos,2i)} &= sin(pos/10000^{2i/d_{model}})\\
    PE_{(pos,2i+1)} &= cos(pos/10000^{2i/d_{model}})
\end{align}

The formulas for sinusoidal positional embeddings can be seen in (4) and (5).   \(d_{model}\) represents the total dimension of the input features, \cite{attention,sasrec,bert4rec}. For the learnable positional embeddings, which is a linear layer without the bias term, the embedding parameters are initialized from \(N(0,1)\). We compared the learnable product embeddings with the product vector of historical behavior in Figure \ref{model}, for making the final prediction.

The use of learnable product embeddings was pivotal in constructing a scalable model, allowing us to compare these embeddings directly with our model's output. This implies that even if a new, unknown product is added to a session, we can still compare our model's output with the existing learnable product vectors to generate a final product recommendation. Consequently, our model exhibits the flexibility to accommodate new and unknown products, enhancing its adaptability and utility in dynamic environments.

All of the input embeddings were concatenated, the dimension of learnable product embeddings is 128, and the  \(d_{model}\)  variable in sinusoidal positional embedding is 128.

\subsection{Style Embeddings}

For extracting the style information, we utilized the neural style transfer algorithm \cite{style}. In the neural style transfer algorithm, the feature maps of a generic object detection algorithm are used for capturing the style information. The filter response of each layer of a convolutional neural network is used to obtain the gram matrices, which are used for calculating the style loss \cite{style}. There are two approaches to how we can transfer style information by using the neural style transfer algorithm.

\begin{figure}[ht]
  \centering
  \includegraphics[width=\linewidth]{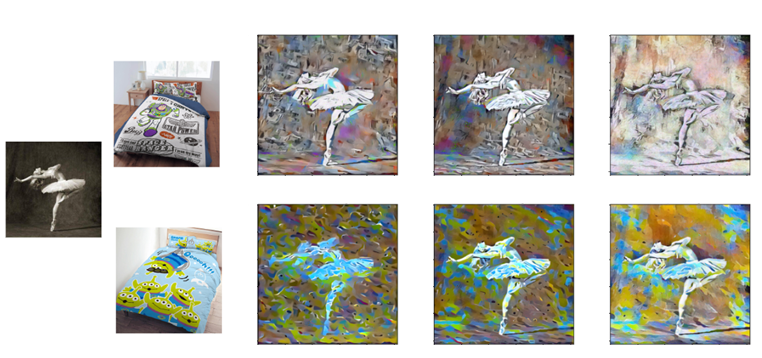}
  \caption{Neural style transfer algorithm updates the content image, with the help of gram matrices, depending on which layers' gram matrices are used and the ratio of 
  \(\frac{L_{style}}{L_{content}}\), different final results can be obtained.} 
\label{style1} 
\end{figure}

Two images are used in the first method of the neural style algorithm, as you can see in Figure \ref{style1}. These images are the content image and the style image. The main idea is to update the content image by using the style image so that the style of the content image becomes similar to the style of the style image. During the training of the neural style transfer algorithm, style and content images are used for calculating the style-loss and the content-loss, respectively, \cite{style}. The gram matrices for both the content and style images are calculated. The neural style transfer algorithm tries to make both gram matrices similar to each other to transfer the style information between the style image and the content image.

\begin{figure}[ht]
  \centering
  \includegraphics[width=180pt]{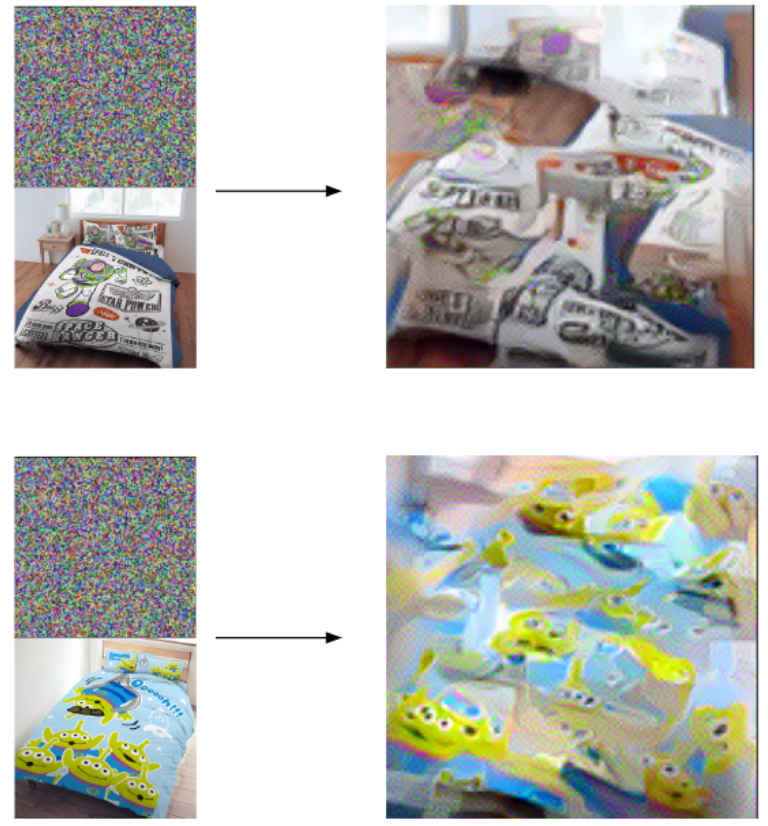}
  \caption{First method of neural style transfer algorithm, algorithm updates the noisy input image with the help of gram matrices.}
\label{style2} 
\end{figure}

\begin{figure*}
\begin{center}
  \includegraphics[width=\textwidth]{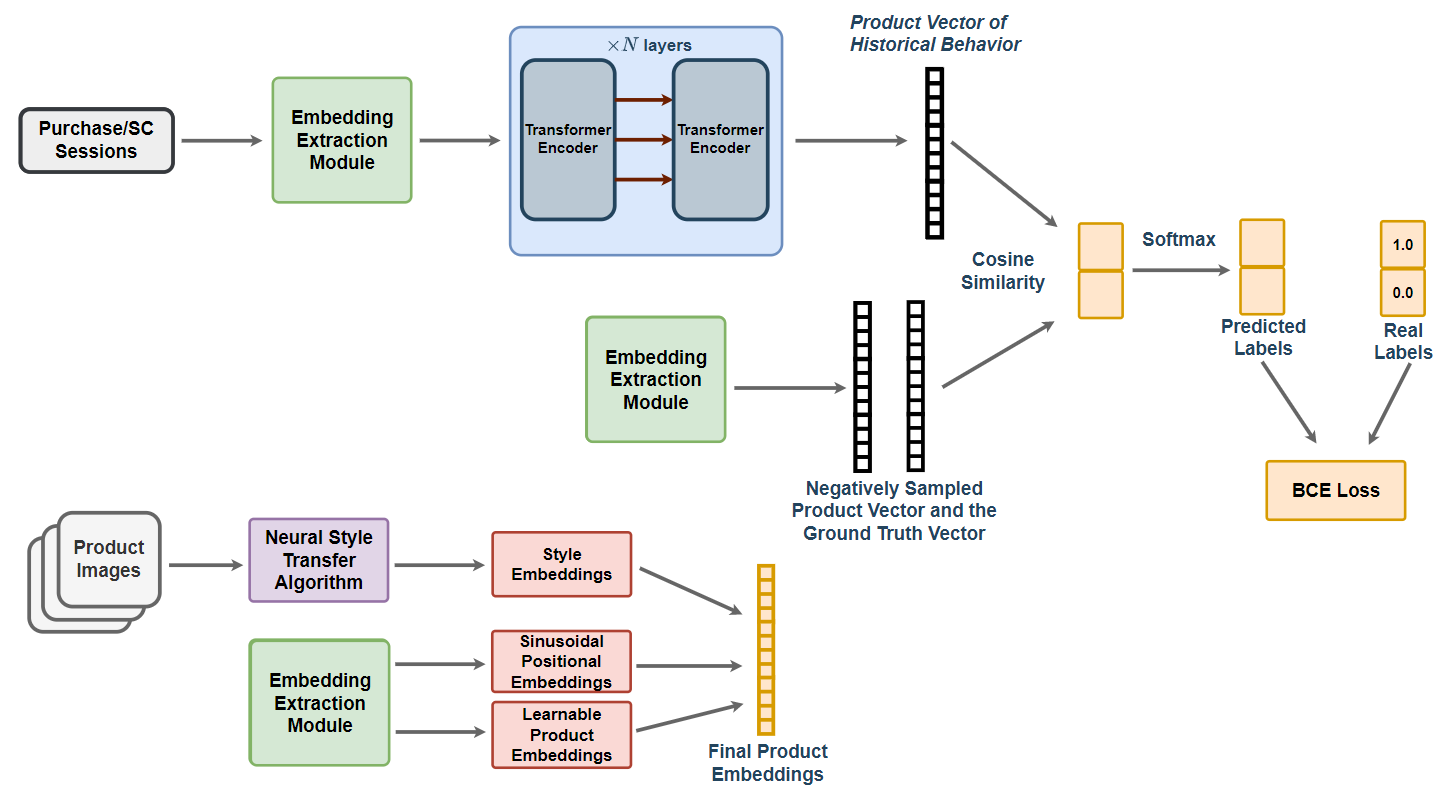}
\end{center}
  \caption{Model Architecture}
\label{model} 
\end{figure*}

In the second method of the neural style transfer algorithm \cite{style}, the style image and the image consisting of Gaussian noise are used, as you can see in Figure \ref{style2}. The main idea is to update the noisy input image so that the content and style of the noisy input image become similar to the style image. The gram matrices for the style and noisy input images are calculated to transfer the style information. The feature maps of the noisy input image and the style image are directly subtracted from each other to transfer the content information. In the experiment shown in Figure \ref{style2}, we transferred only the style information by setting the content loss as zero to visualize style transfer more clearly.

For the style loss, the gram matrices are calculated. Note that both methods utilize the gram matrices since it captures crucial style information. 

\begin{align}
  styleLoss^k &= \frac{1}{4N_k^{2}M_k^{2}} \sum_{} (G_{i}^k - G_{style}^k)^2\\
  L_{style}(input,style) &= \sum_{k=0}^{K}w^{k}styleLoss^k
\end{align}

In formula (6), \(G_{input}\) represents the gram matrix of the noise image, \(G_{style}\) represents the gram matrix of the style image, k represents the layer number, \(N_k\) represents the number of feature maps at layer k, \(M_k\) represents the flattened dimensions of the feature maps. The gram matrices represent the correlation between the feature maps of a specific layer of an object detection algorithm as in formula (6). The gram matrices are calculated by using the dot products of each pair of feature maps in a particular convolutional layer. The gram matrices are obtained separately for multiple network layers, and the respective style losses are calculated. The linear combination of the style losses from different layers of the network is used as the final style loss of the algorithm, \cite{style}. Decreasing the style loss means the style information is transferred between 2 images.

The neural style transfer algorithm can be applied using any object detection algorithm that utilizes convolutional layers. We picked the VGG-19 object detection algorithm \cite{vgg} for extracting the style embeddings from the gram matrices. In our case, we used the gram matrices of the first two layers of the VGG-19 object detection algorithm \cite{vgg} as the style embeddings of the product images. Since there are 64 feature maps in the first and second layers of the VGG-19, the dimension of the style embeddings becomes 64 x 64, and the total dimension of the style embeddings becomes 2 x 64 x 64. Note that the dimension of the gram matrices is independent of the dimension of the feature maps. It depends only on the number of feature maps in a particular convolutional layer. For making the training time of the multi-layer transformer encoder block reasonable, \(O(n^{2}d)\), we applied max-pooling on the gram matrices, so the total dimension of the style embeddings becomes 2 x 16 x 16. Evaluating the effect of gram matrices of other layers is left as future work.

\vspace{-0.2 cm}
\subsection{Shopping Cart Data}


In our dataset, we classify sessions into two types: purchase sessions and shopping cart sessions. Shopping cart sessions, which feature products added to the shopping cart but not ultimately purchased, provide a unique opportunity to enhance the performance of sequential product recommendations. Recognizing the user interest that these items represent, we have devised a strategy to differentiate between these two types of sessions. While both purchase sessions and shopping cart sessions are utilized in the 
training and validation stages, only purchase sessions are included in the testing phase. This approach allows us to more accurately reflect real-world performance, as the primary objective of sequential product recommendation is to predict items users are likely to purchase. Thus, we achieve a realistic gauge of its effectiveness by evaluating the model solely with purchase sessions.

\section{Experiments}

\subsection{Dataset Description}

Our partnering company provided the dataset we used. The data comes from an e-commerce website. The website consists of household goods. The dataset contains pageview, purchase, and shopping cart data of users. We discarded the sessions that consist of only pageviews. In the dataset, there are  490817 interactions (including pageviews, purchase, and shopping cart information) on 38117 user sessions, which can be seen in Table \ref{table:2}. Both purchase and shopping cart sessions contain the same set of products. Shopping cart sessions are longer than purchase sessions in general, the average session length of purchase sessions is 11.24, and the average session of shopping cart sessions is 14.57.\newline

\begin{table}[h!]
\centering
\caption{Statistics of datasets}
\begin{tabular}{p{1.2cm}||p{1.2cm} p{1.2cm} p{1.4cm} p{1.2cm}}
 \hline
 Datasets & \#Sessions & \#Products & Avg.Length & \#Actions \\[1pt] 
 \hline
  Purchase & 19463 & 2991 & 11.24 & 218913 \\
  S.Cart & 18654 & 2991 & 14.57 & 271904 \\ [0.6ex] 
 \hline
\end{tabular}

\label{table:2}
\end{table}

\subsection{Preprocessing}

\begin{figure}[ht]
  \centering
  \includegraphics[width=\linewidth]{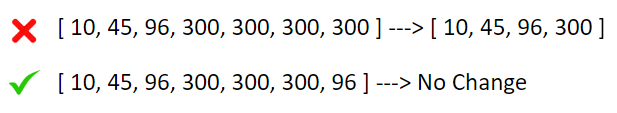}
  \caption{An example of how the repeated final products were removed, each number represents a product.} 
\label{preprocessing} 
\end{figure}

We separated purchase and shopping cart sessions. We removed the overlapped sessions, which contain both purchase and shopping cart products, to see the effect of adding shopping cart data more clearly. We set the max length of the sessions to 20, and we added padding (0) if the sessions were shorter than 20 products. If the sessions were longer than 20 products, we used the last 20 products. We removed the repeated final products as seen in Figure \ref{preprocessing}. We made that change because we know that allowing our model to learn the repeated patterns, could decrease the real-life performance significantly. We wanted our model to learn more complex patterns for better generalization in real-life conditions.\newline\newline\newline

\subsection{Training Procedures}

We split both the purchase data and the shopping cart data  with respect to time. We used the first 14 months of data for training, the next 2 months for validation, and the last 2 months for testing. During training, validation, and testing, we predicted the last items in user sessions, by utilizing the previous items. We used shopping cart sessions only in training and validation but not in testing. We used purchase sessions in training, validation, and testing. We tuned the hidden dimension of the transformer encoder within the range of [8, 16, 32, 64, 128, 256] and the L2 regularization penalty within the range of [0.1, 0.001, 0.0001, 0.00001]. We set the number of transformer blocks and the number of heads as 2, for fair comparison with other benchmarks (BERT4Rec, SASRec). 

For the benchmarks (BERT4Rec, SASRec), we tuned the hyper-parameters according to the descriptions in the respective papers, or we used the recommended parameters. All of the models are reported under their best hyper-parameter settings.

\subsection{Evalaution Methodology}

For our sequential product recommendation system, we design a multi-layer transformer encoder network to extract both the product vector of historical behavior and the learnable product vector as separate components. To make the final prediction, we compare the product vector of historical behavior with the learnable product vectors. To ensure robust evaluation, we employ negative sampling techniques \cite{bpr, chen2017sampling} to select 100 negatively sampled products for each session. The negative samples are combined with the ground truth product, resulting in a set of 101 products. From this set, we sort and select the most relevant products for each session. To assess the performance of our recommendation system, we calculate several evaluation metrics, including Hit Ratio (HR), Mean Reciprocal Rank (MRR), and Normalized Discounted Cumulative Gain (NDCG) \cite{he2017neural, He_2017}. These metrics are calculated for product recommendation list lengths of 5, 10, and 20, providing a comprehensive evaluation of our system's effectiveness in generating accurate and relevant recommendations.

\begin{table*}[htb]
\begin{center}
  \captionsetup{justification=centering}
  \caption{Comparison with the benchmarks. The best results are in boldface, and the second-best results are underlined. Improvement was calculated between \(Style4Rec\) and the best result of BERT4Rec or SASRec.}
  \label{table:benchmarks}
  \begin{tabular}{c||c c c c c c c c c}
  \hline
    Models & HR@5 & HR@10  & HR@20 & NDCG@5 & NDCG@10 & NDCG@20 & MRR@5 & MRR@10 & MMR@20\\
    \hline\hline
    BERT4Rec & 0.677 & \underline{0.756} & \underline{0.834} & 0.583 & 0.609 & \underline{0.629} & 0.552 & 0.563 & 0.568\\
    SASRec & \underline{0.681} & 0.742 & 0.803 & \underline{0.594} & \underline{0.610} & 0.625 & \underline{0.559} & \underline{0.567} & \underline{0.572}\\
    \(Style4Rec\) & \textbf{0.735} & \textbf{0.784} & \textbf{0.838} & \textbf{0.674} & \textbf{0.690} & \textbf{0.704} & \textbf{0.654} & \textbf{0.661} & \textbf{0.665}\\
    \hline
    Improvement & +0.054 & +0.028 & +0.004 & +0.080 & +0.080 & +0.075 & +0.095 & +0.094 & +0.093\\
    \hline
\end{tabular}
\end{center}
\end{table*}

We evaluated the performance of our sequential product recommendation network in 4 different training configurations, which are:
\begin{enumerate}
    \item Purchase Sessions
    \item Purchase Sessions  + Style Embeddings
    \item Purchase Sessions  + Shopping Cart Sessions
    \item Purchase Sessions  + Shopping Cart Sessions + Style Embeddings
\end{enumerate}

By systematically comparing the performance of $4$ different configurations, we thoroughly evaluated each component's individual importance in the sequential product recommendation task. Through this analysis, we were able to identify the optimal combination that yields the best results. This comprehensive approach allows us to understand the relative significance and impact of each part within the recommendation system, leading to valuable insights for improving the overall effectiveness of sequential product recommendations.

\section{Results}

\subsection{Comparison with Baseline Models}


We evaluated the efficiency of our model against two state-of-the-art benchmarks - BERT4Rec and SASRec. We included all the available data -purchase sessions, shopping cart sessions, and style embeddings- for comparing STYLE4Rec with the benchmarks. As you can see in Table \ref{table:benchmarks}, across all recommendation list lengths, our model outshone both BERT4Rec and SASRec in terms of all metrics. The HR@5 metric improved from 0.681 to 0.735, NDCG@5 rose from 0.594 to 0.674, and MRR@5 ascended from 0.559 to 0.654. SASRec obtained the second-best results in 6 evaluation metrics and BERT4Rec obtained the second-best results in 3 evaluation metrics. Considering these results, we can conclude that utilizing style information with the help of neural style transfer algorithms and utilizing shopping cart sessions in training/validation yields significant improvement in sequential product recommendation task.


\subsection{Effect of Style Embeddings and Shopping Cart Data}

As shown in Table \ref{table:ablation}, we tested the model performance on 4 different configurations to evaluate the effect of each method separately. In \(Style4Rec^{1}\) shopping cart sessions and style embeddings were removed. In \(Style4Rec^{2}\) shopping cart sessions were removed. In \(Style4Rec^{3}\) style embeddings were removed. In \(Style4Rec\),  we utilized all the available data, purchase sessions, shopping cart sessions, and style embeddings.

Comparing the \(Style4Rec^{1}\) and \(Style4Rec^{2}\) models in Table \ref{table:ablation}, adding style embeddings to the purchase data improves the model performance on 6 out of 9 metrics, and for MRR@10 the results are the same. We can conclude that extracting style embeddings with the help of the neural style transfer algorithm increases the performance of the sequential product recommendation task. Comparing \(Style4Rec^{1}\) and \(Style4Rec^{3}\) models, adding shopping cart data to purchase data increases the model performance on all of the metrics in all recommendation list lengths. We can conclude that utilizing shopping cart data on training and validation yields meaningful contributions to the sequential product recommendation task. Comparing the \(Style4Rec^{1}\) and \(Style4Rec\) models, adding both shopping cart data and the style embeddings at the same time increases the model performance on all of the metrics in all recommendation list lengths. However, in this case, we evaluated the model performance on a wider model. We increased the number of heads to 8 and the dimension of the learnable product embeddings to 1024 to make the model wider and allow more attention heads to make different predictions. 

The substantial improvements we observed can be attributed to the significant volume of information we incorporated, exceeding all previous instances. Consequently, our model had the capacity to discern more intricate relationships among products, owing to its increased breadth. However, when we endeavored to deepen our model by augmenting the number of transformer blocks, there was no discernible enhancement in the final instance. This can be possibly traced back to the average session length (11.24/14.57) present in our dataset. A deeper model excels at recognizing complex relationships in prolonged sessions. As the average session lengths in our dataset were relatively brief, assessing the model's performance based on a broader model demonstrated superior results, especially in instances when all the data was taken into account.

    

\subsection{Dynamic Recommendation}

\begin{figure}[ht]
  \centering
  \includegraphics[width=\linewidth]{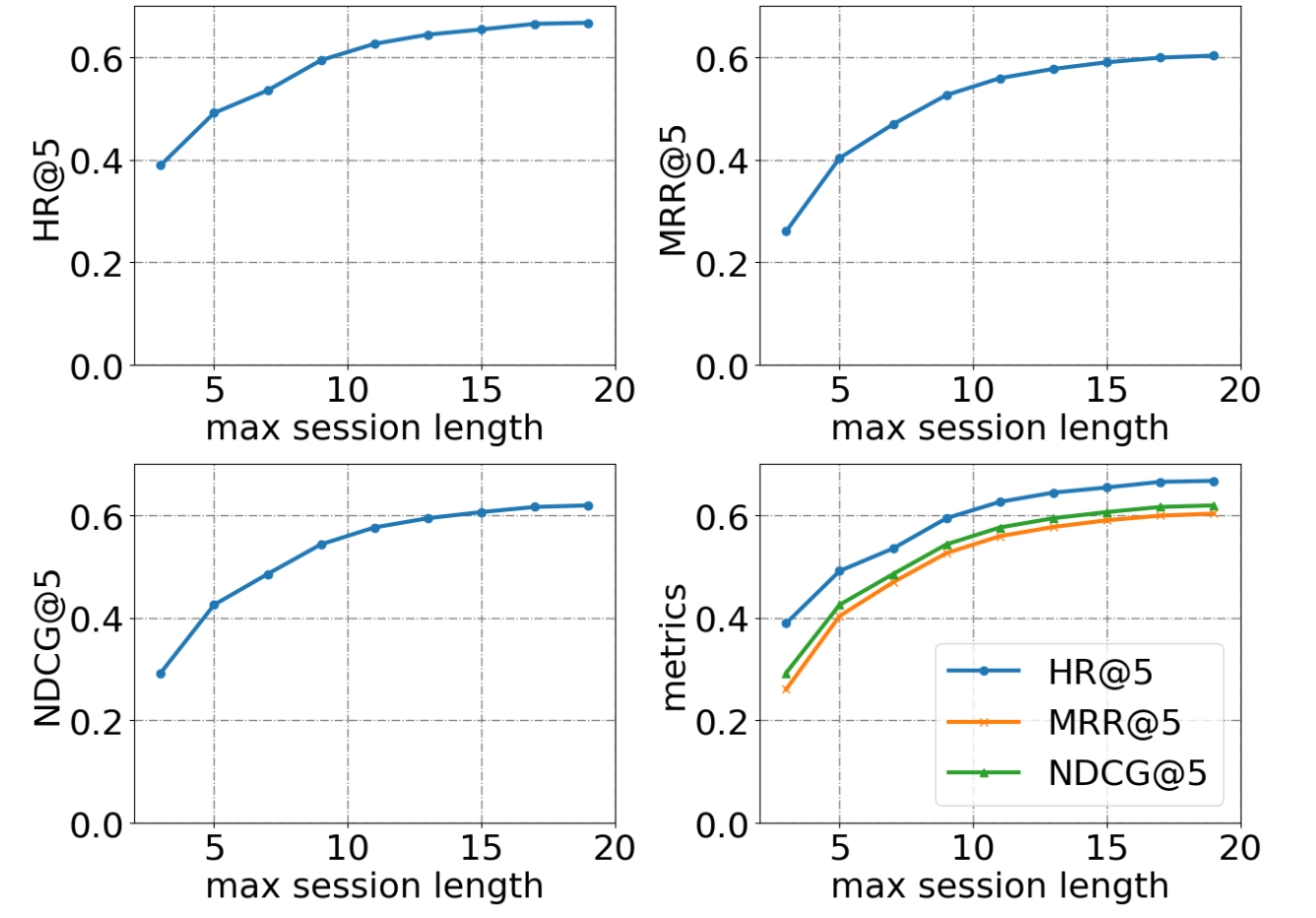}
    \caption{The effect of maximum session length.} 
  \label{sessLength}
\end{figure}
In this section, our aim was to demonstrate the performance of our model under varying maximum session lengths, for evaluating our model's performance in dynamic recommendation task. The dynamic recommendation allows e-commerce websites to recommend products for their users, immediately after those users enter the website. Waiting for users to finish their sessions and then recommending products, might be an ineffective way for e-commerce websites. In prior experiments, we established a uniform maximum session length of 20 across all models and configurations.

As we increased the maximum session length, as in Figure \ref{sessLength}, we observed an increase in all metrics. Increasing the maximum session length allowed our model to utilize the relationships of the products which are far away from the end product. In that way, our model learned more distant and complex relationships between products.

\begin{table*}[htb]
\begin{center}
  \captionsetup{justification=centering}
  \caption{Comparison of the 4 configuration settings of Style4Rec. The best results are in boldface, and the second-best results are underlined. Improvement was calculated between the worst model and the best model, \(Style4Rec^{1}\) and \(Style4Rec\), respectively.}
  \label{table:ablation}
  \begin{tabular}{c||c c c c c c c c c}
    \hline
    Configurations & HR@5 & HR@10  & HR@20 & NDCG@5 & NDCG@10 & NDCG@20 & MRR@5 & MRR@10 & MMR@20\\
    \hline
    \(Style4Rec^{1}\) &  0.670 & 0.718 & 0.769 & 0.622 & 0.637 & 0.650 & 0.606 & 0.612 & 0.616\\
    \(Style4Rec^{2}\) &  0.681 & 0.735 & 0.788 & 0.623 & 0.641 & 0.654 & 0.604 & 0.612 & 0.615\\
    \(Style4Rec^{3}\) &  \underline{0.718} & \underline{0.765} & \underline{0.825} & \underline{0.667} & \underline{0.682} & \underline{0.697} & \underline{0.650} & \underline{0.656} & \underline{0.661}\\
    \(Style4Rec\) & \textbf{0.735} & \textbf{0.784} & \textbf{0.838} & \textbf{0.674} & \textbf{0.690} & \textbf{0.704} & \textbf{0.654} & \textbf{0.661} & \textbf{0.665}\\
    \hline
    Improvement & +0.065 & +0.066 & +0.069 & +0.052 & +0.053 & +0.054 & +0.048 & +0.049 & +0.049\\
    \hline
\end{tabular}
\end{center}
\end{table*}

\begin{table*}[htb]
\begin{center}
  \captionsetup{justification=centering}
  \caption{Effect of negative sampling}
  \label{table:neg}
  \begin{tabular}{c||c c c c c c c c c}
    \hline
    Configurations & HR@5 & HR@10  & HR@20 & NDCG@5 & NDCG@10 & NDCG@20 & MRR@5 & MRR@10 & MMR@20\\
    \hline\hline
    \(Style4Rec_{neg}\) &  0.670 & 0.718 & 0.769 & 0.622 & 0.637 & 0.650 & 0.606 & 0.612 & 0.616\\
    \(Style4Rec_{w\backslash o \ neg}\) &  0.505 & 0.549 & 0.585 & 0.434 & 0.448 & 0.457 & 0.410 & 0.416 & 0.419\\
    \hline
    Decrease & -0.165 & -0.169 & -0.184 & -0.188 & -0.189 & -0.193 & -0.196 & -0.196 & -0.197\\
    \hline
\end{tabular}
\end{center}
\end{table*}

We also observed that increasing the maximum session length to more than 18 does not significantly improve performance, since the distant products might have less effect on the final decision-making.

\subsection{Effect of Negative Sampling}

Negative sampling was employed as our primary strategy for assessing model performance. However, in real-world recommendation systems, the necessity arises to rank all available products to formulate the final product recommendations. As you can see in Table \ref{table:neg}, we removed the negative sampling process in an attempt to simulate our model's performance under realistic conditions. This modification resulted in a decline across all metrics, with decreases ranging between 0.165 and 0.197. With negative sampling, the prediction scope was confined to 101 products; however, in its absence, the prediction space expanded dramatically to encompass 2991 products. The decrease in prediction space using negative sampling generally leads to improved results, as it reduces the complexity of the task at hand. The disparity between laboratory conditions and actual market complexities underlines the need for continuous innovation and model refinement to ensure these advanced algorithms can reliably deliver optimal recommendations in practice.

\section{Conclusion}

We have introduced a multi-layer transformer encoder network that integrates the neural style transfer algorithm to incorporate style information. Alongside this, we devised a training methodology that emphasizes the distinctions between purchased and shopping cart products. Our comprehensive experiments demonstrated a significant performance enhancement in sequential product recommendation tasks when style embeddings and shopping cart data were incorporated into the transformer recommender network. Furthermore, our model was shown to surpass
existing state-of-the-art benchmarks across multiple evaluation metrics.
Notably, our model was designed with scalability in mind. By storing learnable product vectors, we have enabled the comparison of a new product's vector with existing ones, allowing for an adaptable and continuously evolving recommender system.
However, while we have made significant strides in improving recommendation quality, it is essential to acknowledge that there is always room for further refinement. Our future work will be geared toward assessing the performance of our model in scalable tasks. The challenge lies in maintaining and even improving upon this level of performance as we scale to handle larger datasets and more complex recommendation scenarios. However, we are confident that with further research and continuous development, our model will continue to evolve and excel in the ever-changing landscape of product recommendation systems.\newline\newline\newline\newline

\bibliography{aaai25}
\end{document}